\documentclass[11pt]{article}
\usepackage{epsfig,rotate}
\usepackage{a4}
\usepackage{amssymb}
\newcommand{\ASSERGILNGS} {1} 
\newcommand{\BARI}        {2}
\newcommand{\BERN}        {3}
\newcommand{\BOLOGNA}     {4}
\newcommand{\LAQUILA}     {5}
\newcommand{\LYON}        {6}
\newcommand{\NAPOLI}      {7}
\newcommand{\NEUCHATEL}   {8}
\newcommand{\PADOVA}     {10}
\newcommand{\ROMA}       {11}
\newcommand{\SALERNO}    {12}

\newcommand{\OperaInstitutes}{
\ASSERGILNGS . Laboratori Nazionali del Gran Sasso dell'INFN, 67010 Assergi (L'Aquila), Italy \\
\BARI        . Dipartimento di Fisica dell'Universit\`a  di Bari and INFN, 70126 Bari, Italy \\
\BERN        . University of Bern, CH-3012 Bern, Switzerland \\
\BOLOGNA     . Dipartimento di Fisica dell'Universit\`a  di Bologna and INFN, 40127 Bologna, Italy \\
\LAQUILA     . Dipartimento di Fisica dell'Universit\`a  dell'Aquila and INFN, 67100 L'Aquila, Italy\\
\LYON        . IPNL, Universit\'e Claude Bernard Lyon 1, CNRS/IN2P3, 69622 Villeurbanne, France\\
\NAPOLI      . Dipartimento di Fisica dell'Universit\`a Federico II di Napoli and INFN, 80125 Napoli, Italy \\
\NEUCHATEL   . Universit\'e de Neuch\^atel, CH-2000 Neuch\^atel, Switzerland\\
\PADOVA      . Dipartimento di Fisica dell'Universit\`a  di Padova and INFN, 35131 Padova, Italy \\
\ROMA        . Dipartimento di Fisica dell'Universit\`a  di Roma ``La Sapienza" and INFN, 00185 Roma, Italy \\
\SALERNO     . Dipartimento di Fisica dell'Universit\`a  di Salerno and INFN, 84084 Fisciano, Salerno, Italy \\
}

\newcommand{\OperaAuthorList}{
L.~Arrabito$^{\LYON}$,
D.~Autiero$^{\LYON}$, 
C.~Bozza$^{\SALERNO}$,
S.~Buontempo$^{\NAPOLI}$,
Y.~Caffari$^{\LYON}$,
L.~Consiglio$^{\BOLOGNA}$,
M.~Cozzi$^{\BOLOGNA}$,
N.~D'Ambrosio$^{\ASSERGILNGS}$,
G.~De~Lellis$^{\NAPOLI}$,
M.~De~Serio$^{\BARI}$,
F.~Di~Capua$^{\NAPOLI}$,
D.~Di~Ferdinando$^{\BOLOGNA}$,
N.~Di~Marco$^{\LAQUILA}$,
A.~Ereditato$^{\BERN}$,
L.~S.~Esposito$^{\ASSERGILNGS}$,
S.~Gagnebin$^{\NEUCHATEL}$,
G.~Giacomelli$^{\BOLOGNA}$,
M.~Giorgini$^{\BOLOGNA}$,
G.~Grella$^{\SALERNO}$,
M.~Hauger$^{\NEUCHATEL}$,
M.~Ieva$^{\BARI}$,
J.~Janicsko Csathy$^{\NEUCHATEL}$,
F.~Juget$^{\NEUCHATEL}$,
I.~Kreslo$^{\BERN}$,
I.~Laktineh$^{\LYON}$,
A.~Longhin$^{\PADOVA}$,
G.~Mandrioli$^{\BOLOGNA}$,
A.~Marotta$^{\NAPOLI}$,
J.~Marteau$^{\LYON}$,
P.~Migliozzi$^{\NAPOLI}$,
P.~Monacelli$^{\LAQUILA}$,
U.~Moser$^{\BERN}$,
M.~T.~Muciaccia$^{\BARI}$,
A.~Pastore$^{\BARI}$,
L.~Patrizii$^{\BOLOGNA}$,
C.~Pistillo$^{\BERN}$,
M.~Pozzato$^{\BOLOGNA}$,
G.~Romano$^{\SALERNO}$,
G.~Rosa$^{\ROMA}$,
A.~Russo$^{\NAPOLI}$,
N.~Savvinov$^{\BERN}$,
A.~Schembri$^{\ROMA}$,
L.~Scotto~Lavina$^{\NAPOLI}$,
S.~Simone$^{\BARI}$,
M.~Sioli$^{\BOLOGNA}$,
C.~Sirignano$^{\SALERNO}$,
G.~Sirri$^{\BOLOGNA}$,
P.~Strolin$^{\NAPOLI}$,
V.~Tioukov$^{\NAPOLI}$.\\
}
\begin{document}

\title{\bf Electron/pion separation with an Emulsion Cloud Chamber by using a Neural Network}

\maketitle

\author{\noindent \\ \OperaAuthorList }

\begin{flushleft}
\footnotesize{\OperaInstitutes }
\end{flushleft}

\baselineskip=14pt

\vspace{0.2cm}

{\bf{Abstract}}
We have studied the performance of a new algorithm for electron/pion separation in an Emulsion Cloud Chamber (ECC) made of lead and nuclear emulsion films. The software for separation consists of two parts: a shower reconstruction algorithm and a Neural Network that assigns to each reconstructed shower the probability to be an electron or a pion. The performance has been studied for the ECC of the OPERA experiment \cite{unknown:2006ki}. 

The $e/\pi$ separation algorithm has been optimized by using a detailed Monte Carlo simulation of the ECC and tested on real data taken at CERN (pion beams) and at DESY (electron beams). The algorithm allows to achieve a 90\% electron identification efficiency with a pion misidentification smaller than 1\% for energies higher than 2 GeV.  


\section{Introduction}

The Emulsion Cloud Chamber \cite{kaplon,emulsion2} consists of a modular structure made of a sandwich of passive material plates interleaved with emulsion films. It combines the high-precision tracking capabilities of nuclear emulsions and the large mass achievable by employing passive material as a target. Among past applications the ECC was successfully used in the DONUT experiment for the first direct observation of the tau-neutrino \cite{Kodama:2000mp}. By assembling a large quantity of ECC modules, it is possible to realize $\mathcal{O}(kton)$ fine-grained vertex detector optimized for the study of $\nu_\tau$ appearance. 

The ECC has been adopted by the OPERA Collaboration \cite{unknown:2006ki,opera} for a long-baseline search of $\nu_{\mu}\rightarrow\nu_{\tau}$ oscillations. OPERA is designed to obtain a unambiguous signature (observation of $\nu_\tau$ appearance) of $\nu_{\mu} \rightarrow \nu_{\tau}$ oscillations in the parameter region indicated by atmospheric neutrino experiments  \cite{sk,kam,macro,soudan2} and confirmed by long-baseline experiments with accelerator neutrinos: K2K \cite{k2k} in Japan and MINOS \cite{minos} in the USA. The detector is located in the underground Gran Sasso Laboratory. It exploits the Cern to Gran Sasso (CNGS) beam, optimised for $\nu_{\tau}$ appearance, with a baseline of 730 km \cite{tappearance}. OPERA is a hybrid experiment based on the use of  ECC and of electronic detectors for a rough location of the events in the ECC and for full event reconstruction. The basic OPERA ECC module has dimensions of 12.7$\times$10.2$\times$7.5 cm$^3$; it consists of a sequence of 56 lead plates (1 mm thick) and 57 emulsion films (44 $\mu$m thick emulsion layers on either side of a 205 $\mu$m thick plastic base \cite{emulsion1}). The total length of an OPERA ECC module is about 10 $X_0$.

In addition to the decay detection of short-lived particles \cite{Kodama:2002dk}, the ECC allows momentum measurement by Multiple Coulomb Scattering \cite{Lellis:2003xt} and the particle identification through ionization measurement \cite{Toshito:2004tc}. 

The high granularity of the emulsions also allows an excellent electromagnetic shower identification, hence the separation of electrons and pions. This is obtained by exploiting their different behavior in passing through and interacting in an ECC. In the 1-10 GeV energy range, of interest for OPERA, electrons loose energy essentially by bremsstrahlung and charged pions mainly by ionization. This motivates two complementary approaches to identify electrons and pions:

\begin{itemize}
\item an electron quickly develops an electromagnetic shower in lead (critical energy $\sim 10$ MeV). The total number of tracks, as well as the different longitudinal and transverse profiles of the showers, can be then used for particle identification;
\item  going through a material, the energy remains almost constant for pions whilst strongly decreases for electrons. Therefore, Multiple Coulomb Scattering presents different longitudinal profiles for electrons and pions, that a $\chi^2$-based separator may be built. A method based on this approach is possible with an ECC and is described in \cite{ejapan}, where it is shown that an electron efficiency of 90\% with a pion contamination of 6\% is achievable by using 56 emulsion films interleaved with lead plates. Another study \cite{thmichela} shows that with this method it is possible to achieve an electron efficiency above 90\% for a pion contamination not lower than 5\% by using 30 emulsion films.
\end{itemize}

In this paper we follow the first approach by using a new algorithm for shower reconstruction and a new algorithm based on a Neural Network (NN) for $e/\pi$ separation. A preliminary study to exploit a Neural Network for this purpose was presented in \cite{luillo}.

\section{Data taking with test-beams}
\label{expdata}
Experimental data have been collected in an electron beam at DESY and in a pion beam at CERN.

The ECC exposures to pions took place in the CERN PS-T7 beam-line with beams of 2, 4 and 6 GeV. The electronic detector setup consisted of a Cherenkov counter, two scintillation counters, two multi-wire chambers and a lead-glass centered on the beam line. In order to reduce the electron contamination, a 2.5 cm thick lead preshower was added upstream of the last focusing magnet\footnote{Given the beam line configuration, it was not possible to locate the preshower before the last bending magnet, where it would have been most efficient. Anyhow, a significative reduction of the electron contamination in the beam was achieved.} of the T7 beam line and upstream from the Cherenkov counter. The electron contamination was measured by using the combined information of the Cherenkov and lead-glass detectors. It amounted to $\sim$0.6\% at 2 GeV and was less than 0.4\% at 4 and 6 GeV. The muon contamination was measured by analyzing passing through tracks in the ECC under test ($\sim$38\% at 2 GeV, $\sim$6\% at 4 and 6 GeV) \cite{thmichela}. In order to have reasonable statistics, the ECCs integrated a beam intensity corresponding to about 2000 tracks/cm$^2$. The incident angle of the beam with respect to the ECC emulsion films was 50 mrad.

Electron data were taken at DESY T24 beam-line, where a sub-Hz rate per cm$^2$ was achieved in the energy range from 1 to 6 GeV. The sub-Hz rate was mandatory in order to achieve a small particle density in the ECC, as needed to avoid overlaps of electromagnetic showers. The electronic detector setup consisted of three scintillation counters used as a trigger, a multi-wire chamber to measure the beam profile and a lead-glass calorimeter for electron energy measurement. Two ECCs, both with 20 emulsion films, were exposed to electrons: one at 6 GeV high density (100 particles/cm$^2$) beam and one at 3 GeV low density (1 particle/cm$^2$) beam. A third ECC was brought to DESY but not exposed, in order to estimate the background accumulated from ECC production to the emulsion development. The low energy contamination of the beam, due to the interactions of the electrons with the beam pipe and the beam-line elements, was estimated to be 5.7\% at 1 GeV and 2.3\% at 6 GeV.

The emulsion films have been scanned by using the European Scanning System (ESS) developed for the high-speed automatic scanning of the OPERA emulsion films\footnote{A scanning system for the high-speed automatic scanning of the emulsion films (called SUTS) has been developed also in Japan \cite{suts}}. The main features and performance of the ESS are presented in \cite{mic3}. High speed particle tracking for the ESS is described in \cite{scan1}, precision measurements in \cite{scan2}, alignments with cosmic ray muons in \cite{cosmic} and event analysis in \cite{fedra}.

\section{The shower reconstruction}
\label{shower}

The algorithm for the reconstruction of the shower follows an iterative procedure. For each base-track\footnote{A base-track is obtained connecting through the plastic base two segments (micro-tracks) reconstructed in each emulsion layer. For details on the micro-track and base-track reconstruction we refer to \cite{scan1}.} (in the following called "selector") base-tracks matching it in the downstream films are searched for. The matching criteria are based on angular and position requirements. The angular displacement $\delta\theta$ is defined as the angle difference between the selector and the base-track candidate; the position displacement $\delta r$ is the transverse distance between the selector and the candidate extrapolated back to the selector. Any matched candidate becomes a selector and so on. In order to take into account the reconstruction efficiency, a base-track candidate is allowed to be extrapolated back at most for 3 films, then it is discarded. Monte Carlo simulations have shown that an adequate background rejection is achieved if one sets $\delta\theta  < 50$ mrad and $\delta r < 100$ $\mu$m\,.
 
\begin{figure}[htpb]
\begin{center}
\leavevmode
\epsfig{figure=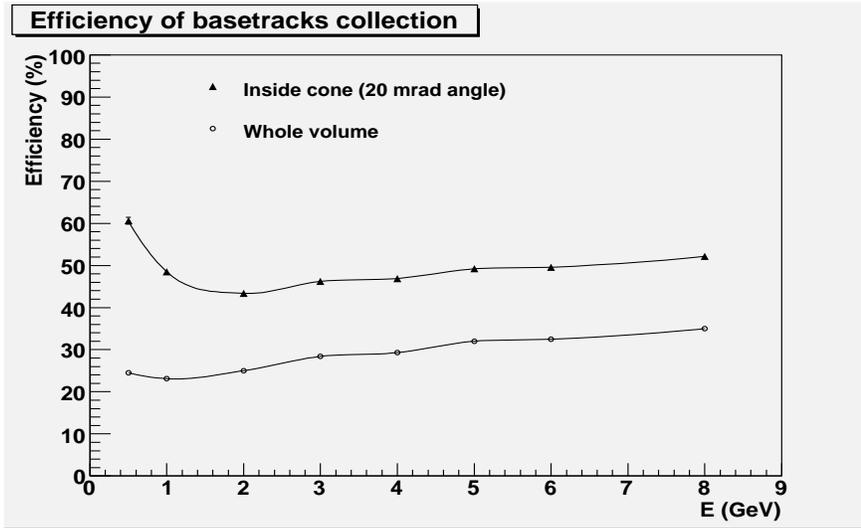,height=7.cm,width=11.5cm}
\caption{ Efficiency for base-tracks collection versus the electron energy. The two curves refer to different event samples: the total number of base-tracks in the whole volume and inside the fiducial volume, respectively.}
\label{effbt}
\end{center}
\end{figure}

\begin{figure}[htpb]
\begin{center}
\leavevmode
\epsfig{figure=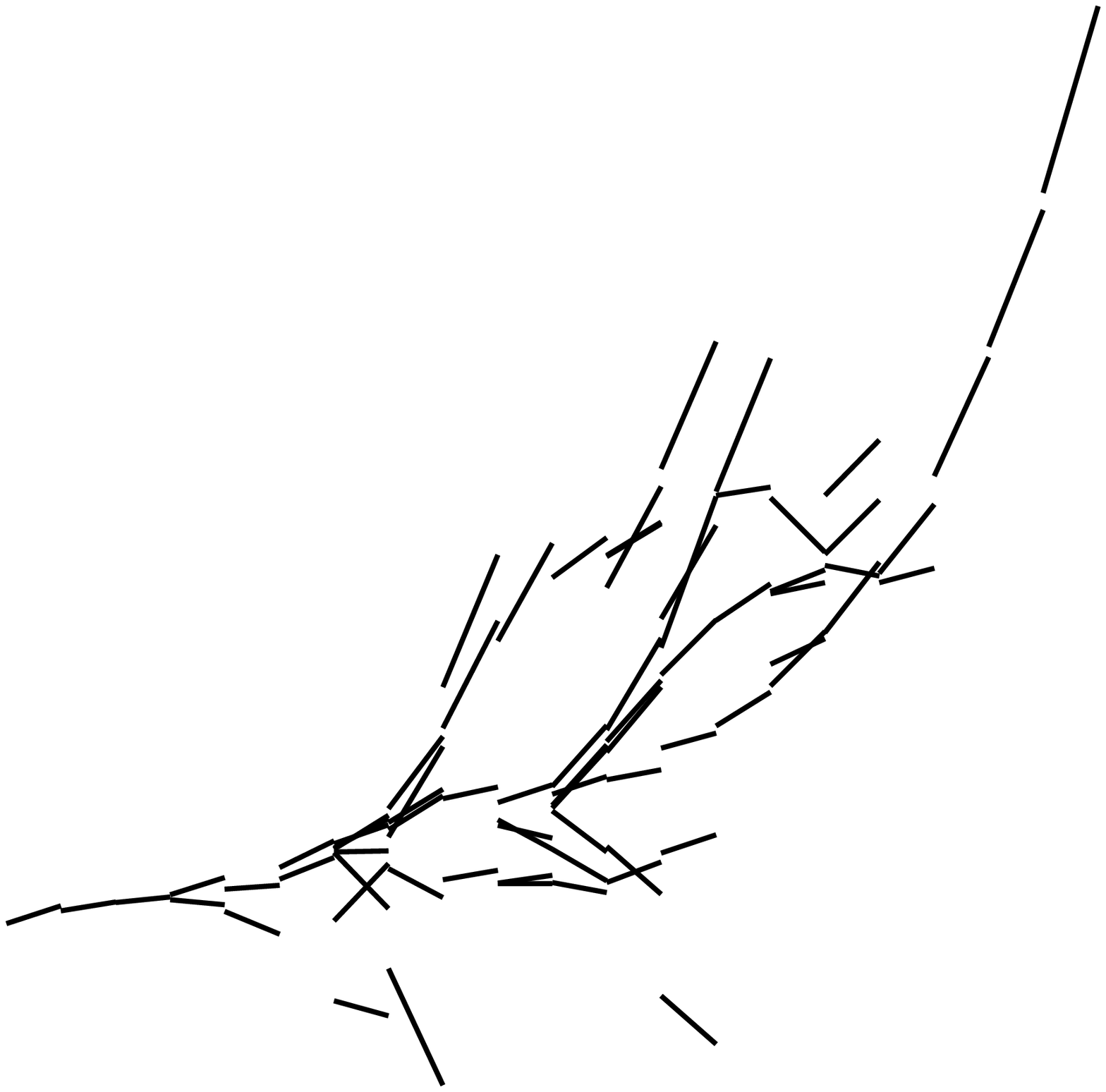,height=6.cm,width=6.5cm}
\epsfig{figure=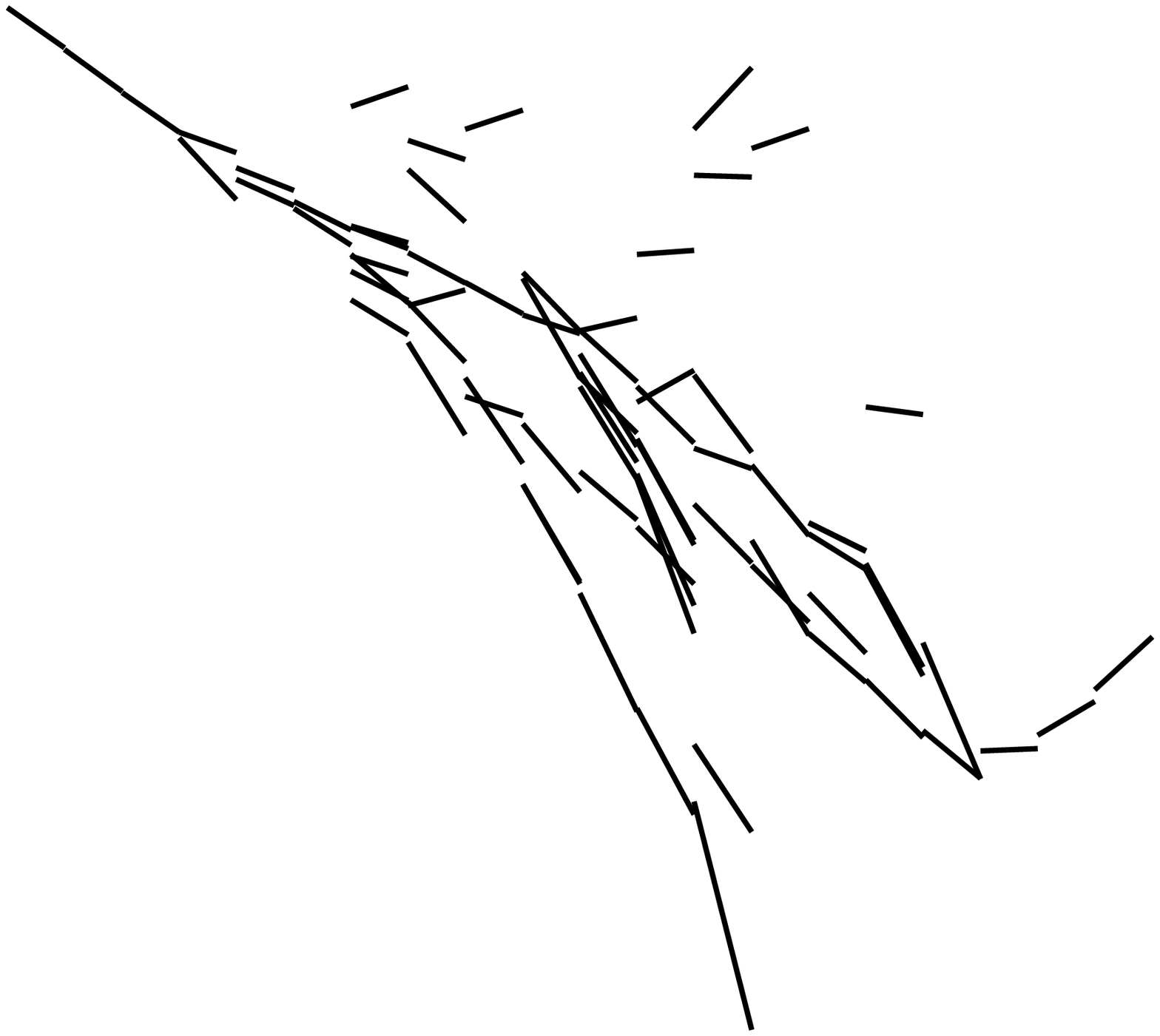,height=6.cm,width=6.5cm}
\caption{ xz projection (left) and yz projection (right) of a reconstructed shower generated by a 6 GeV electron interacting in the $\sim$3.3 $X_0$ECC exposed at DESY. Each segment corresponds to a base-track associated with the reconstructed electromagnetic shower.}
\label{6GeV_1}
\end{center}
\end{figure}

In order to minimize the background, a fiducial volume cut around the shower axis is applied. Base-tracks must be within a cone (with axis defined by the slope of the first base-track belonging to the shower) with an opening angle of 20 mrad. This angle has been optimized by a Monte Carlo simulation where each track has been digitized by using the tools described in \cite{geant3,orfeo}. In addition, in order to minimize showers' overlaps, one requires that the radius of the cone does not exceed 400 $\mu$m. Therefore, beyond 15 emulsion films the cone becomes a cylinder with 400 $\mu$m radius. To further reduce the background, one removes base-tracks previously assigned to a track starting outside the fiducial volume. Finally, one imposes that showers traverse at least 4 emulsion films. The efficiency to collect base-tracks belonging to the shower is shown in Fig. \ref{effbt}. A reconstructed 6 GeV electron shower is shown in Fig. \ref{6GeV_1}.

The performance of the algorithm has been tested with experimental data and tuned with a detailed Monte Carlo simulation \cite{geant3,orfeo}. The main sources of inefficiency come from the scanning of the emulsion films and from the requirement on the minimum length of the shower ($> 3$ films). The efficiency of the shower reconstruction algorithm as a function of the energy is given in Table \ref{tableineff}. The errors given in the table and in the following ones are statistical only. The probability for a particle to be correctly identified is given by the product of the above efficiencies and of the ones given by the NN, discussed in the following Section.

\begin{table}[htbp]
\begin{center}
\begin{tabular}{|c|c|c|}   \hline
E & Pion & Electron\\ 
(GeV)& \% & \%  \\ \hline \hline 
0.5  &69.0$\pm$1.6 & 46.2$\pm$1.8\\
1  &87.8$\pm$1.1 & 78.9$\pm$1.5\\
2  &88.5$\pm$1.1 & 91.0$\pm$1.0\\
3  &89.4$\pm$1.0 & 91.6$\pm$1.0\\
4  &91.8$\pm$0.9 & 93.0$\pm$0.9\\
5  &91.8$\pm$0.9 & 94.6$\pm$0.8\\
6  &91.6$\pm$0.9 & 95.1$\pm$0.8\\
8  &90.6$\pm$0.9 & 95.5$\pm$0.8\\
\hline
\end{tabular}
\caption{Efficiency of the shower reconstruction algorithm for simulated pions and electrons, with background added.}
\label{tableineff}
\end{center}
\end{table}

%

\section{The Neural Network}
\label{TheNN}

\begin{figure}[htpb]
\begin{center}
\leavevmode
\epsfig{figure=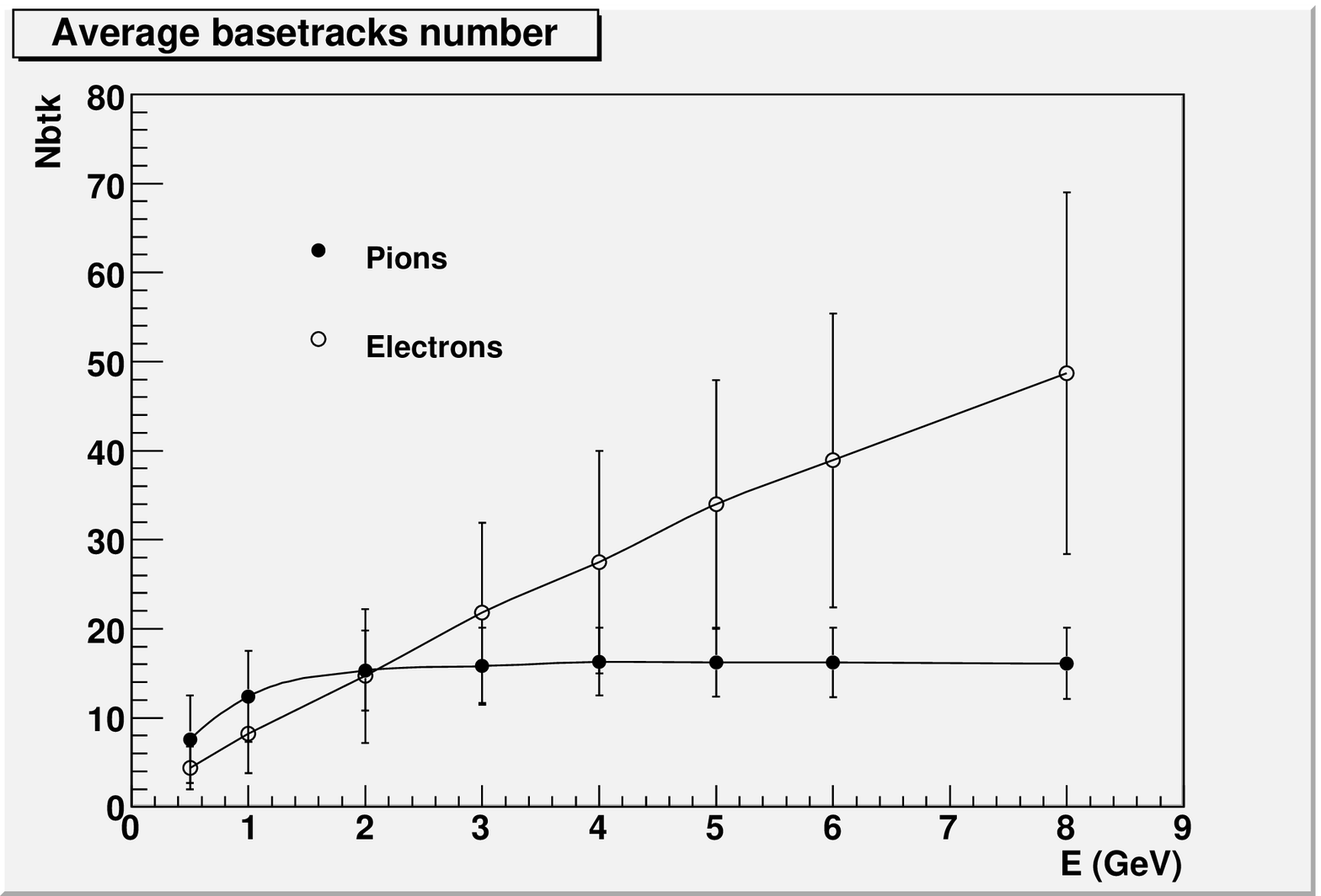,height=6cm,width=9.cm}
\epsfig{figure=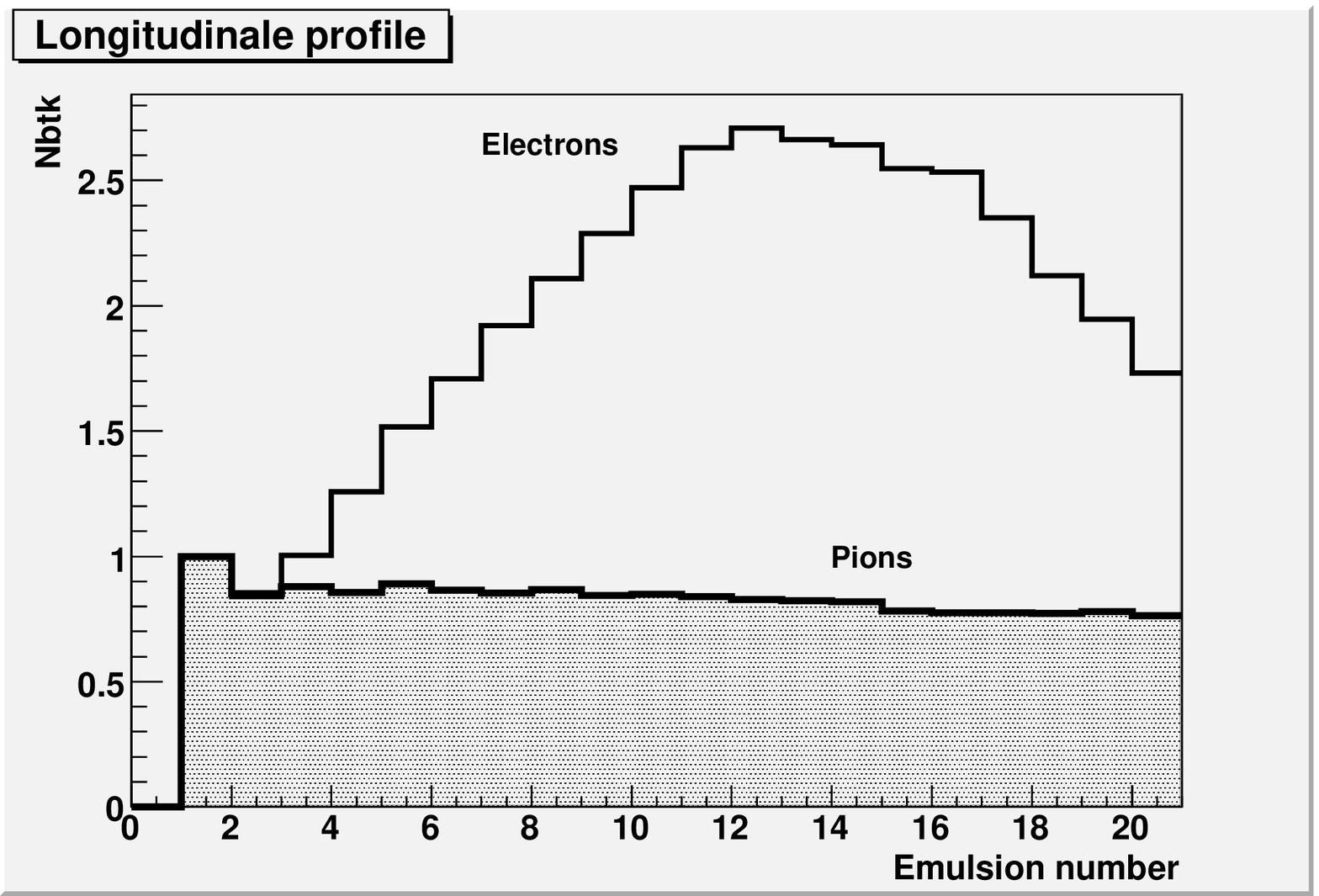,height=6cm,width=9.cm}

\caption{Top: average number of base-tracks in reconstructed showers versus energy. The error bars reflect the fluctuation on the number of base-tracks produced into a shower. Bottom: mean longitudinal shower profile for 6 GeV particles. Both plots refer to an ECC with 20 emulsion films interleaved with 19 lead plates. }
\label{profile}
\end{center}
\end{figure}

\begin{figure}[htpb]
\begin{center}
\leavevmode
\epsfig{figure=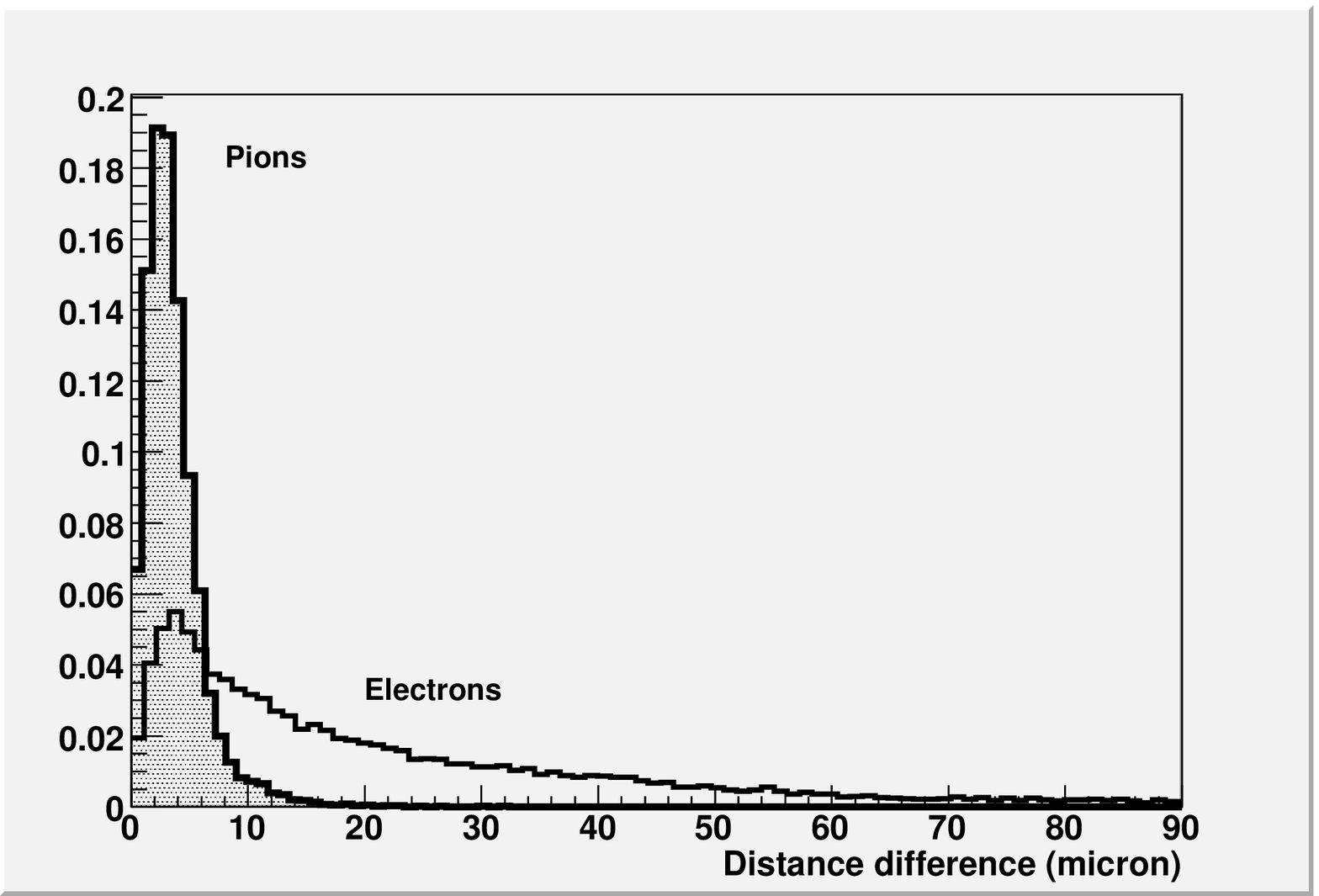,height=6.cm,width=9.cm}
\epsfig{figure=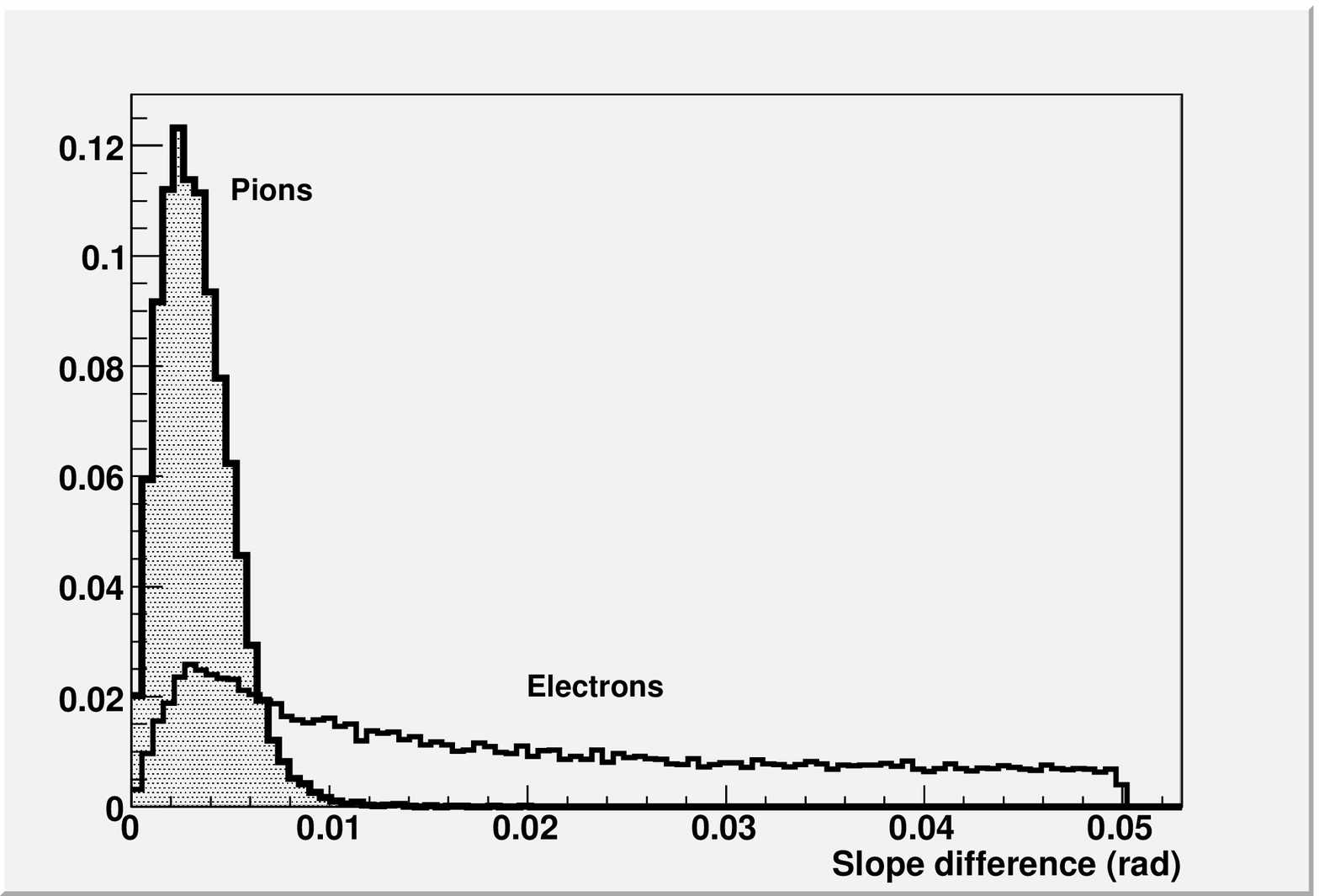,height=6.cm,width=9.cm}
\caption{Distribution of  $\delta r$ (top) and $\delta\theta$ (bottom) for electrons and pions of 6 GeV.}
\label{delta}
\end{center}
\end{figure}

Particle identification is performed through an algorithm based on a Neural Network \cite{mlp}. Each reconstructed shower is fed into the NN. The longitudinal profile and the number of base-tracks (shown in Fig. \ref{profile}) as well as the $\delta r$  and  $\delta\theta$ distributions (shown in  Fig. \ref{delta}) are very different for electron and pions. They are used as inputs for the NN.

The ECC exposed to the DESY beam had 20 emulsion films, together with the lead plates corresponding to $\sim$3.3 $X_0$. We will present in Section \ref{smaller} a Monte Carlo study if the ESS performance as a function of the number of films, i. e. the traversed thickness in terms of radiation lengths. With 20 emulsion films to reconstruct the shower the NN has 23 input variables, defined as follows: 
\begin{itemize}
\item 1 variable corresponding to the number of base-tracks in the reconstructed shower ($n_{btk}$) (Fig. \ref{profile} top);
\item  18 variables describing the longitudinal profile (Fig. \ref{profile} bottom). The first two bins are removed since they are very similar for pions and electrons;
\item 2 variables corresponding to the mean and the RMS of the  $\delta r$ distribution (Fig. \ref{delta} top);
\item 2 variables corresponding to the mean and the RMS of the $\delta\theta$ distribution (Fig. \ref{delta} bottom).
\end{itemize}
Besides the 23 input neurons, the NN has two hidden layers with 63 and 21 neurons, and one output neuron. The training is stopped after 120 "epochs" when the predefined ``sampling error'', computed on the validation sample, reaches a plateau before starting to increase.\\

\section{Results of the electron/pion separation algorithm}

The electron efficiency $\epsilon_{e \rightarrow e}$ and the pion contamination $\eta_{\pi \rightarrow e}$ are defined, respectively, as: 
\begin{center}
$\epsilon_{e \rightarrow e} \equiv \frac{n_{e \rightarrow e}}{N_e} \ \ \ \ \ \ \ \ \ \ \ \ \ \ \ \ \eta_{\pi \rightarrow e} \equiv \frac{n_{\pi \rightarrow e}}{N_{\pi}}$   
\end{center}
where $n_{e \rightarrow e}$ ($n_{\pi \rightarrow e}$) is the number of electrons (pions) classified as electrons by the NN and $N_e$ ($N_{\pi}$) is the total number of true electrons (pions) sent to the NN. The pion efficiency $\epsilon_{\pi \rightarrow \pi}$ and the electron contamination $\eta_{e \rightarrow \pi}$ are similarly defined. 

\begin{figure}[htpb]
\begin{center}
\leavevmode
\epsfig{figure=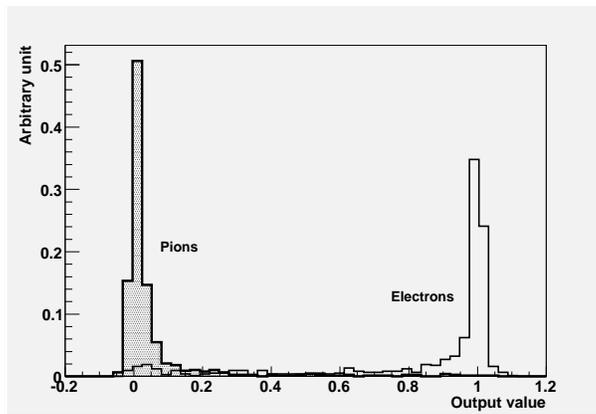,height=5.5cm,width=8.cm}
\caption{ Output value given by the neural network for simulated 2 GeV pions and electrons.}
\label{output}
\end{center}
\end{figure}

The particle identification is performed by applying a cut on the output neuron value, that ranges between 0 and 1 (Fig. \ref{output}). The actual cut depends on the analysis needs. For example, for some analyses a high electron identification efficiency is required, to a certain extent regardless the pion misidentification $\eta_{\pi \rightarrow e}$. For other applications a small $\eta_{\pi \rightarrow e}$ is specially important. Here, we study two selections. One demands a high electron efficiency ($\epsilon_{e \rightarrow e} > 80\%$) regardless the pion misidentification (from now on Selection A). The other demands low pion misidentification ($\eta_{\pi \rightarrow e}<1\%$) regardless the electron identification efficiency (from now on Selection B). Given both the small number of tracks associated to an electromagnetic shower and the large contribution from the pion charge exchange process in the low energy range ($< 2$ GeV), this is the most difficult region where to achieve simultaneously a good $\epsilon_{e \rightarrow e}$ and a low $\eta_{\pi \rightarrow e}$.  The value of the cut is imposed at 1 GeV and applied at all energies. The results shown in this paper can be further improved by applying an energy dependent cut. The study of the measurement of the electromagnetic shower energy is in progress and will be the subject of a forthcoming publication.

\subsection{Ideal case of absence of background}
\label{nobck}
In this Section we present the results on the electron identification efficiency and on the pion contamination obtained with simulated events in the absence of background. Only pure pion and electron events are generated and reconstructed with the shower algorithm presented in Section \ref{shower}.

\begin{figure}[htbp]
\begin{center}
\leavevmode
\epsfig{figure=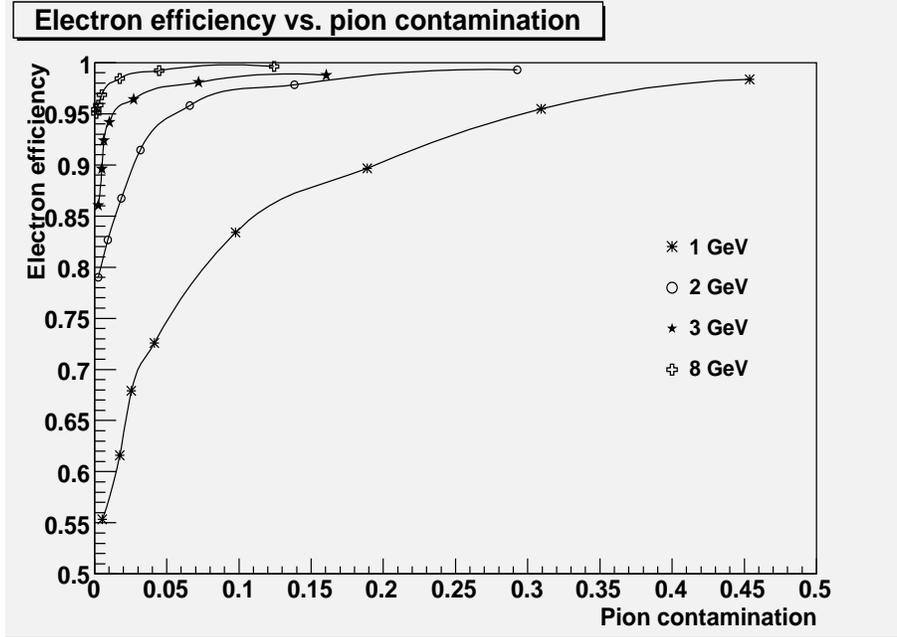,height=8.5cm,width=12.cm}
\caption{ Monte Carlo simulation of the electron efficiency versus pion contamination for different energies by using 20 emulsion films.}
\label{effvscont}
\end{center}
\end{figure}

The NN has been trained using $\sim 14500$ electron and $\sim 16500$ pion events with a flat energy spectrum in the range 0.5 to 6 GeV. The validation sample, different than the training sample, contained about 800 electrons and 800 pions with energies of 0.5, 1, 2, 3, 4, 5, 6, 8 GeV. The results of the validation sample re shown in Fig. \ref{effvscont}. Table \ref{table1} shows that if Selection A is tuned at 1 GeV, an electron identification efficiency larger than 80\% can be obtained over the whole energy range with a contamination from pion misindentification of about 1\% for energies above 2 GeV. $\eta_{\pi \rightarrow e}$ is much higher for lower energies. If Selection B is tuned at 1 GeV, $\eta_{\pi \rightarrow e}$ is below 1\% for energies above 1 GeV and an electron identification efficiency lower than 80\% for energies $\leq$ 2 GeV is obtained. In general, for energies lower than 1 GeV it is very difficult to reach a small pion misidentification, because of the pion charge exchange reaction and to the relatively small number of base-tracks in the electromagnetic showers.

\begin{table}[htpb]
\begin{center}
\begin{tabular}{|c|c|c||c|c|}   \hline
E & $\epsilon_{e \rightarrow e}$ &  $\eta_{\pi \rightarrow e}$& $\epsilon_{e \rightarrow e}$ &  $\eta_{\pi \rightarrow e}$\\ 
(GeV)& \% & \% & \%  & \% \\ \hline \hline 
 &  \multicolumn{2}{|c||}{Selection A} &\multicolumn{2}{|c|}{Selection B }    \\ \hline
0.5  &90.1$\pm$1.5 & 55.3$\pm$2.0& 52.9$\pm$2.6  &18.3$\pm$1.6\\
1 &81.7$\pm$1.6 & 9.0$\pm$1.0&58.8$\pm$1.9  &1.0$\pm$0.4\\
2 &89.2$\pm$1.1 & 1.7$\pm$0.5&80.5$\pm$1.5  &0.6$\pm$0.3 \\
3 &93.5$\pm$0.9 & 0.7$\pm$0.3&90.1$\pm$1.1  &0.1$\pm$0.1  \\
4  &95.6$\pm$0.8 & 0.5$\pm$0.2&92.4$\pm$1.0  &0.1$\pm$0.1 \\
5  &98.4$\pm$0.5 & 0.5$\pm$0.2&96.5$\pm$0.7  &0.2$\pm$0.2  \\
6  &96.8$\pm$0.6 & 0.4$\pm$0.2&94.4$\pm$0.8  &0.1$\pm$0.1\\
8   &98.7$\pm$0.4 & 0.6$\pm$0.3&97.7$\pm$0.5  &0.2$\pm$0.2\\
  \hline
\end{tabular}
\caption{Electron efficiency and pion contamination for pure simulation using 20 emulsion films ($\sim$3.3 $X_0$). The  output NN value is fixed at 0.62 for selection A and 0.82 for selection B.}
\label{table1}
\end{center}
\end{table}

\subsection{Test-beam data and comparison with simulations}
\label{bckadd} 

In order to train the NN under conditions similar to the test-beam exposure, we added some background to the simulated sample. The background accumulated in the emulsions has been obtained by scanning emulsion films not exposed to the beams and has been added by software to simulated events. 

We have evaluated the background that is accepted by the shower algorithm. We found that the fraction of fake base-tracks associated to a shower does not exceed 5\%. In order to study the effect of the background on the performance of the NN, we applied the same cut on the output neuron value as in Section \ref{nobck}. The results are shown in Tables  \ref{table2A} and \ref{table2B} for selections A and B, respectively. By comparing the results for simulated events  given in Tables \ref{table2A} and \ref{table2B} with those given in Table \ref{table1}, one sees that the electron efficiency is worsened by about 5\% with respect to the case without background. The pion to electron misidentification is mildly affected  (except for energies below 1 GeV) by the presence of background.

\begin{table}[htpb]
\begin{center}
\begin{tabular}{|c|c|c||c|c|}   \hline 
E & $\epsilon_{e \rightarrow e}$ & $\eta_{\pi \rightarrow e}$& $\epsilon_{e \rightarrow e}$ & $\eta_{\pi \rightarrow e}$\\
(GeV)& \% & \% & \%  & \% \\ \hline \hline 
 &  \multicolumn{2}{|c||}{Simulated Events} &\multicolumn{2}{|c|}{Data }    \\ \hline
0.5  &90.5$\pm$1.5 & 81.3$\pm$1.6& ---  & ---\\
1 &81.1$\pm$1.6 & 8.4$\pm$1.0& --- & ---\\
2 &82.0$\pm$1.4 & 0.7$\pm$0.3& ---  &2.0$\pm$0.3 \\
3 &89.2$\pm$1.2 & 0.5$\pm$0.3&$>61$  & ---  \\
4  &89.9$\pm$1.1 & 0.4$\pm$0.2& --- &0.4$\pm$0.1 \\
5  &93.9$\pm$0.9 & 0.4$\pm$0.2& --- & ---  \\
6  &93.1$\pm$0.9 & 0.4$\pm$0.2&96.3$\pm$0.8  &0.4$\pm$0.2\\
8   &95.3$\pm$0.8 & 0.1$\pm$0.1& --- & ---\\
\hline
\end{tabular}
\caption{Electron efficiency and pion contamination for simulated and test-beam data assuming Selection A. The  output NN value is fixed at 0.62.}
\label{table2A}
\end{center}
\end{table}

\begin{table}[htpb]
\begin{center}
\begin{tabular}{|c|c|c||c|c|}   \hline
E & $\epsilon_{e \rightarrow e}$ & $\eta_{\pi \rightarrow e}$& $\epsilon_{e \rightarrow e}$ & $\eta_{\pi \rightarrow e}$\\ 
(GeV)& \% & \% & \%  & \% \\ \hline \hline 
 &  \multicolumn{2}{|c||}{Simulated Events} &\multicolumn{2}{|c|}{Data }    \\ \hline
0.5  &53.4$\pm$2.6 & 28.9$\pm$1.9& ---  & ---\\
1 &59.4$\pm$2.0 & 0.9$\pm$0.4& --- & ---\\
2 &73.6$\pm$1.6 & 0.1$\pm$0.1& ---  &0.5$\pm$0.1 \\
3 &82.2$\pm$1.4 & 0.1$\pm$0.1&80$\pm$18  & ---  \\
4  &86.0$\pm$1.3 & 0.1$\pm$0.1& --- &0.3$\pm$0.1 \\
5  &90.3$\pm$1.1 & 0.1$\pm$0.1& --- & ---  \\
6  &90.2$\pm$1.1 & 0.2$\pm$0.2&94.7$\pm$0.9 &0.2$\pm$0.1\\
8   &94.1$\pm$0.8 & 0.1$\pm$0.1& --- & ---\\
\hline
\end{tabular}
\caption{Electron efficiency and pion contamination for simulated and test-beam data assuming Selection B. The  output NN value is fixed at 0.82.}
\label{table2B}
\end{center}
\end{table}

The NN trained with the background superimposed to the simulated data has then been applied to the test-beam data described in Section \ref{expdata}. For the ECC with 3 GeV low-density electrons (1 electron/cm$^2$) an area containing only 5 electrons, identified by the known angle of the beam with respect to emulsion films, has been measured. For the ECC with 6 GeV  high-density electrons (100 electrons/cm$^2$), the number of electrons contained in the scanned area is estimated to be about 670. Among those, 620 electrons satisfy the shower reconstruction criteria (Section \ref{shower}) and are fed in the NN. For  2, 4, 6 GeV pions, 2747, 2548, 1591 events, respectively, have been fed into the NN. They have been selected by using the known angle of the beam with respect to the emulsion films.

The results obtained with real data are summarized in Tables \ref{table2A} and \ref{table2B} for Selection A and B, respectively. One can see that the results with test-beam data agree reasonably well with the simulation. 
Note that the pion beam exploited during the ECC exposures had an intrinsic electron contamination of about 0.5\%, see Section \ref{expdata}. The lower limit at 95\% C.L. shown in Table \ref{table2A} has been computed  assuming a binomial distribution \cite{D'Agostini:1995fv}. We computed a lower limit since all electron-beam tracks (5 in total) have been correctly identified.

\subsection{$e$/$\pi$ separation as a function of the traversed $X_0$}
\label{smaller}
Having checked the reliability of the Monte Carlo simulation, as described in the previous Section, we have studied $\epsilon_{e \rightarrow e}$ and $\eta_{\pi \rightarrow e}$ as a function of the number of traversed emulsions films in the energy range 0.5 to 8 GeV. Events have been simulated for 15 ($\sim 2.5 X_0$), 30 ($\sim 5.0 X_0$) and 50 ($\sim 8.3 X_0$) emulsion films interleaved with 1 mm thick lead plates. The results are shown in Tables \ref{table3A} and \ref{table3B} for Selection A and B, respectively. The upper limits (95\% C.L.) have been computed  assuming a binomial distribution \cite{D'Agostini:1995fv}. We have computed upper limits since none of the pion-beam tracks (800 in total) has been misidentified as an electron.

Going from 15 to 30 emulsion films, the electron identification and the pion misidentification improve both for Selection A and Selection B. However, adding more films slightly worsen the performance. In fact, adding more films the fiducial volume, as defined in Section \ref{shower}, increases and the signal/background ratio decreases. An improvement of the performance could be obtained by using in the analysis an energy dependent number of films.

\begin{table}[htpb]
\begin{center}
\begin{tabular}{|c|c|c||c|c||c|c|}   \hline
E & $\epsilon_{e \rightarrow e}$ & $\eta_{\pi \rightarrow e}$& $\epsilon_{e \rightarrow e}$ & $\eta_{\pi \rightarrow e}$& $\epsilon_{e \rightarrow e}$ & $\eta_{\pi \rightarrow e}$\\ 
(GeV)&\% & \% & \%&\% &\% &\%\\ \hline \hline 
 &   \multicolumn{2}{|c||}{50 films} & \multicolumn{2}{|c||}{30 films} &\multicolumn{2}{|c|}{15 films }    \\ \hline
0.5  &92.7$\pm$ 1.4 & 76.5 $\pm$ 1.8 &92.7$\pm$1.4  &78.7$\pm$1.7&83.5$\pm$2.0 & 62.9$\pm$2.0\\
1 &79.9$\pm$1.6 & 14.3 $\pm$ 1.8 & 80.2$\pm$1.6  &15.8$\pm$1.3&80.8$\pm$1.6 & 6.4$\pm$0.9\\
2 &84.3$\pm$ 1.3 & 2.2 $\pm$ 0.5 &85.0$\pm$1.3  &2.5$\pm$0.6 &78.3$\pm$1.5 & 0.9$\pm$0.3\\
3 &90.3$\pm$ 1.1 & 0.9$\pm$ 0.3&91.4$\pm$1.0  & 0.9$\pm$0.3 &82.0$\pm$1.4 & 0.4$\pm$0.2 \\
4  &91.2$\pm$1.1 & 1.1 $\pm$ 0.4&91.6$\pm$1.0  &0.9$\pm$0.3 &86.9$\pm$1.3 & 0.5$\pm$0.3\\
5  &93.3 $\pm$0.9 &1.8 $\pm$0.5&95.0$\pm$0.8  &1.6$\pm$0.4 &88.6$\pm$1.2 &0.2$\pm$1.2 \\
6  &92.8$\pm$ 1.0& 0.6 $\pm$0.3&95.0$\pm$0.8  &0.9$\pm$0.3&88.8$\pm$1.2 & 0.2$\pm$0.2\\
8   &93.2$\pm$ 0.9 & 0.7 $\pm$0.3&95.7$\pm$0.7  &0.6$\pm$0.3&92.0$\pm$1.0 & 0.4$\pm$0.2\\
\hline
\end{tabular}
\caption{Electron efficiency and pion contamination for Selection A with the simulated events with added background using 50, 30 and 15  emulsion films. The output value is fixed at 0.58, 0.58 and 0.73, respectively.}
\label{table3A}
\end{center}
\end{table}

\begin{table}[htpb]
\begin{center}
\begin{tabular}{|c|c|c||c|c||c|c|}   \hline
E & $\epsilon_{e \rightarrow e}$ & $\eta_{\pi \rightarrow e}$& $\epsilon_{e \rightarrow e}$ & $\eta_{\pi \rightarrow e}$& $\epsilon_{e \rightarrow e}$ & $\eta_{\pi \rightarrow e}$\\ 
(GeV)& \% & \% & \%&\% &\% &\%\\ \hline \hline 
 &   \multicolumn{2}{|c||}{50 films} & \multicolumn{2}{|c||}{30 films} &\multicolumn{2}{|c|}{15 films }    \\ \hline
0.5  &28.5 $\pm$2.4 & 16.0$\pm$1.5 &31.8$\pm$2.5  &19.2$\pm$1.6&45.3$\pm$2.6 & 15.3$\pm$1.5\\
1 &40.4$\pm$ 2.0 & 1.2 $\pm$0.4 &43.5$\pm$2.0  &1.1$\pm$0.4&56.8$\pm$2.0 & 0.9$\pm$0.4\\
2 &63.4$\pm$1.8 & 0.4$\pm$ 0.2&64.8$\pm$1.8  &0.1$\pm$0.1 &66.1$\pm$1.7 & 0.3$\pm$0.2\\
3 &78.4$\pm$1.5 & 0.1$\pm$0.1&78.2$\pm$1.5  & 0.1$\pm$0.1 &75.2$\pm$1.6 & 0.1$\pm$0.1 \\
4  &82.5$\pm$ 1.4& 0.1 $\pm$0.1 &84.1$\pm$1.4  &0.1$\pm$0.1 &75.2$\pm$1.6 & 0.1$\pm$0.1\\
5  &86.8 $\pm$1.2 &0.2$\pm$0.2&87.6$\pm$1.2  &0.5$\pm$0.2 &80.8$\pm$1.5 & $<0.4$ \\
6  &81.6$\pm$ 1.4& 0.1$\pm$0.1&87.1$\pm$1.2  &0.2$\pm$0.2&83.4$\pm$1.4 & 0.2$\pm$0.2\\
8   &80.8 $\pm$1.4 & $<0.4$&91.9$\pm$1.0  &$<0.4$ &89.0$\pm$1.1 & $<0.4$\\
\hline
\end{tabular}
\caption{Electron efficiency and pion contamination for Selection B with the simulated events with added background using 50, 30 and 15  emulsion films. The output value is fixed at 0.93, 0.93 and 0.88, respectively.}
\label{table3B}
\end{center}
\end{table}

\subsection{Impact of a cosmic-ray exposure on the $e$/$\pi$ separation}
During the OPERA running, each ECC tagged as a candidate to host a neutrino interaction is extracted from the target and, before its unpacking for the development of the emulsion films, exposed to cosmic-rays at a facility outside the underground hall. ECCs are exposed in a site shielded  by a 40 cm iron slab from the cosmic-ray electromagnetic component. With this configuration, cosmic-ray muons hit the ECC with a rate of $\sim$2 muons/mm$^2$/day inside a 400 mrad cone with respect to the vertical direction. This provides reference tracks useful to intercalibrate and align the emulsion films \cite{cosmic}. 
Cosmic muons have an average momentum of $\sim$4 GeV, and about 70\% of them have a momentum larger than 1 GeV.
Some high energy muons produce bremsstrahlung photons and initiate an electromagnetic cascade. We have studied the impact of this cosmic-ray background on the $e$/$\pi$ separation.

\begin{figure}[htpb]
\begin{center}
\leavevmode
\epsfig{figure=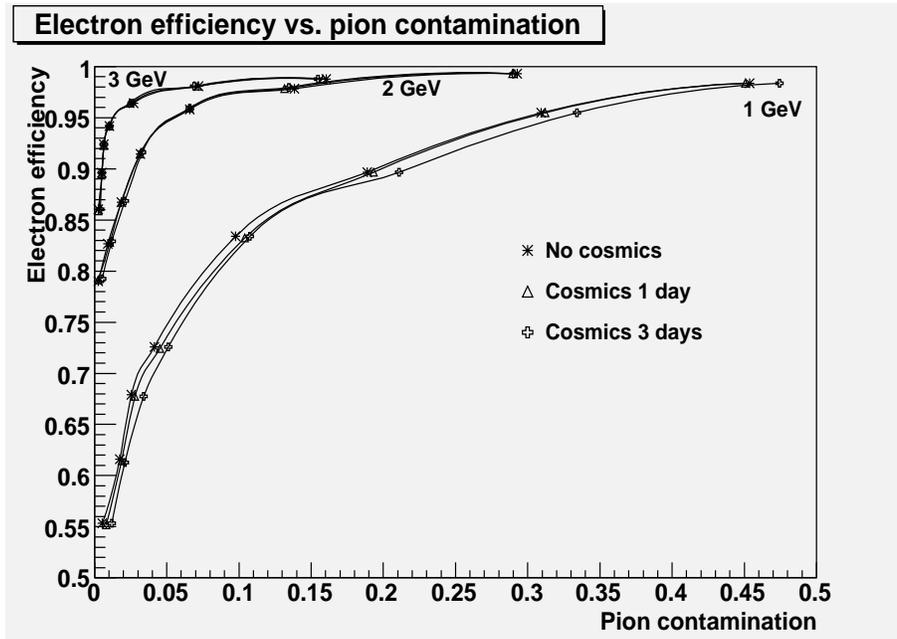,height=8.5cm,width=12.cm}
\caption{ Monte Carlo electron efficiency vs pion contamination with different cosmic-ray exposure time, using 20 emulsion films.}
\label{effvscont_cosmics}
\end{center}
\end{figure}

The cosmic-ray exposure conditions and their interaction products have been simulated and the muons traced inside
the ECC down to 1 MeV kinetic energy. 
We have simulated the background accumulated in 1, 2 and 3 days and superimposed it to the simulated data presented in Section \ref{bckadd}. No effect on the $e$/$\pi$ separation is observed for energies $\geq$ 2 GeV, as shown in Fig. \ref{effvscont_cosmics}. A slight worsening of the performance is observed only at 1 GeV.

\section{Conclusion}

We present the performance of a new algorithm for $e/\pi$ separation in an Emulsion Cloud Chamber (ECC) made of emulsion films interleaved with lead plates. The algorithm consists of two parts: a shower reconstruction algorithm and a Neural Network that assigns to each reconstructed shower the probability to be an electron or a pion. The performance have been studied for the ECC of the OPERA experiment in the CNGS beam. 

We show that the shower reconstruction algorithm has an efficiency higher than 90\% for energies above 1 GeV. We have shown that, depending on the requirements of the analysis, it is possible to achieve either high electron identification efficiency (more than 80\%) or small pion misidentification (smaller than 1\%). 

This study is relevant for the search of $\nu_\mu\rightarrow\nu_\tau$ and $\nu_\mu\rightarrow\nu_e$ oscillations by the OPERA experiment in the CNGS. Given the flexibility of the algorithm it can be easily adapted to OPERA analyses, whose requirements may be conflicting. As an example, the $\nu_\mu\rightarrow\nu_e$ oscillations analysis requires very low pion misidentification \cite{numunue}, while the rejection of the background from $\nu_e$ and $\bar{\nu}_e$ induced charm production requires high efficiency electron identification.

We also studied the impact of the exposure of an lead/emulsion ECC to cosmic-rays as required for film alignment and intercalibration in OPERA. It was shown that after 3 days of cosmic-ray exposure the electron to pion separation deteriorates only very slightly.

 \section*{Acknowledgements}
We acknowledge the cooperation of the members of the
OPERA Collaboration and we thank many colleagues for
discussions and suggestions. We gratefully acknowledge the
invaluable support of the technical staff in our laboratories;
in particular we thank M. Di Marino, V. Di Pinto, F. Fiorello, M. Hess, P. Pecchi, A. Ruggieri, H.-U. Sch\"uetz, V. Togo and C. Valieri for their contributions. We warmly acknowledge support from our funding agencies.
We thank INFN also for providing
fellowships and grants (FAI) for non Italian citizens.

\end{document}